\documentclass[a4paper,10pt]{article}
\usepackage[utf8]{inputenc}
\usepackage[T1]{fontenc}
\usepackage[amssymb]{SIunits}
\usepackage{amsmath, amsfonts, amssymb, amstext, amsbsy}
\usepackage{txfonts, pxfonts, pifont}
\usepackage{fancyhdr}
\usepackage{alltt} 
\usepackage{indentfirst}
\usepackage{vmargin}   
\usepackage{graphicx, color, float}
\usepackage{algorithmic, algorithm}
\usepackage[normal]{subfigure} 
\usepackage[normalem]{ulem}
\usepackage{marvosym, latexsym, wasysym}
\usepackage{verbatim}
\usepackage{booktabs} 
\usepackage[small, bf]{caption}
\usepackage{textcomp}
\usepackage{mathrsfs}
\usepackage[affil-it]{authblk}
\usepackage{hyperref}
\hypersetup{colorlinks,
  debug=False,
  linkcolor=blue,  %%% cor do tableofcontents, \ref, \footnote, etc
  citecolor=red,  %%% cor do \cite
  urlcolor=green,   %%% cor do \url e \href
  bookmarksopen=true,
  pdftitle={Enhancement of the nonlinear optical absorption of the E7 liquid crystal at the nematic-isotropic transition}}
\title{Enhancement of the nonlinear optical absorption of the E7 liquid crystal at the nematic-isotropic transition}

\author{S.L. G\'omez\thanks{Electronic address: \texttt{sgomez@uepg.br}; Corresponding author}}
\author{V.M. Lenart}
\affil{Departamento de F\'isica, Universidade Estadual de Ponta
Grossa, PR, Brazil\\
Tel.: +55-42-32203779\\
Fax: +55-42-32203042\\}

\author{I.H. Bechtold}\affil{Departamento
de F\'isica, Universidade Federal de Santa Catarina, SC, Brazil.}

\author{A.M. Figueiredo Neto}
\affil{Instituto de F\'isica, Universidade de S\~ao Paulo,
S\~ao Paulo, SP, Brazil.}

\date{Dated: \today \\ ~\\ \Large{This paper will be published in Volume 42 of the Brazilian Journal of Physics.}
}

\begin{document}

\maketitle

\begin{abstract}
  We present an experimental study of the nonlinear optical absorption
  of the eutectic mixture E7 at the nematic-isotropic phase transition
  by the Z-scan technique, under continuos-wave excitation at
  $532$\,nm.  In the nematic region, the effective nonlinear optical
  coefficient $\beta$, which vanishes in the isotropic phase, is
  negative for an extraordinary beam and positive for an ordinary
  beam. The parameter $S_{NL}$, whose definition in terms of the nonlinear
  absorption coefficient follows the definition of the
  optical-order parameter in terms of the linear dichroic ratio, behaves like
  an order parameter with critical exponent $0.22 \pm 0.05$, in good
  agreement with the tricritical hypothesis for the N-I transition.
\end{abstract}

\section{Introduction}

After numerous theoretical and experimental studies, the
nematic-isotropic (N-I) transition in liquid crystals still poses
certain controversies and reveals new, interesting facets. For
example, although geometrical arguments lead to the conclusion that the
nematic$\leftrightarrows$isotropic transition is of first-order
\cite{Degennes}, measurements of enthalpy and volume changes at the
transition, when compared with experimental results at melting, fully
support the scenario of a weak first-order phase transition with
strong fluctuations, which may lead to an effective critical behavior
and an effective critical exponent
\cite{Keyes,Filippini,Drozd,Jadzyn,Mcgraw}.  Although there are no
systematic or conclusive studies on this thermodynamic behavior, a few
experimental studies point to the occurrence of tricritical,
rather than just critical effective exponents \cite{Mukherjee,Longa}.

Curiously, there is scarce literature on the behavior of the nonlinear
optical response of liquid crystals at this phase transition. Recently
we reported on the divergence of the nonlinear optical
birrefringence $\Delta n_{2}$ in the nematic side of the N-I
transition of the thermotropic liquid crystal E7, measured with a cw
laser at $\lambda=532\,\nano\meter$ \cite{Lenart}. The critical
exponent seems to corroborate the tri-critical character of this phase
transition.  Previously, Li et al.~\cite{Li} had shown that the
optical absorption of a pure sample of 5CB under ms-pulses of a
$\lambda=514\,\nano\meter$ cw laser is linear at the nematic side of
the N-I transition, i. e., $\left(\beta_{\parallel,\perp}=0\right)$.
Although multiphoton absorption processes are usually observed under
intense pulsed laser irradiation, there are many reports on the
observation of two-photon absorption under irradiation of middle or
low power cw lasers in highly absorbing dye-doped liquid crystal, as
consequence of a two-photon absorption
process \cite{Catalano,Fischer,Ara,Murazawa} or to reorientation of the
nematic director \cite{Tabiryan}. 

The main purpose of this paper is to report on the observation of
nonlinear optical absorption in a planar oriented sample of the
thermotropic liquid crystal E7 and its enhancement in the neighborhood
of the N-I phase transition, under irradiation by a middle-power
$\lambda=532\,\nano\meter$ cw laser. In section 2 we describe the
Z-scan technique, which is used for measuring the nonlinear optical
absorption coefficient $\beta$. In section 3 we present the
experimental data and discuss the possibles mechanisms for the
nonlinear absrption observed in our case. Finally, in section 4 we
sum up the conclusions.

\section{Experimental details}

The abbreviation E7 stands for a liquid crystal mixture consisting
of several types of cyanobiphenyls, mainly 5CB, and in less quantity,
triphenyl. It exhibits a nematic phase in the temperature interval
from $-10\,\celsius$ up to the transition to the isotropic phase
at $T_{NI}=58\,\celsius$. In this experiment we used commercial
E7 (Merck), without additional purification. The samples were conditioned
in a parallel glass cell, separated by $20\,\micro\meter$ thick spacers.
The glass plates were coated with PVA (polyvinyl alcohol) and buffed
for homogeneous planar alignment of the liquid crystal. The sample
was placed in a hot stage (INSTEC) on a computer-controlled translational
stage (Newport). The temperature of the sample was controlled with
a precision of $0.2\,\celsius$ between $20\,\celsius$ and $T_{NI}$.

The Z-scan technique exploits the formation of a lens in a medium
by a tightly focused Gaussian-profile laser beam, measuring the transmitted
intensity of the sample as a function of position along the direction
of propagation of the beam (the $z$-axis) \cite{Sheik}. The magnitude
and sign of the nonlinear absorption $\beta$ can be obtained by means
of the open-aperture configuration of the $Z$-scan technique. A sketch
of the experimental setup is shown in Fig. 1. In this setup the totality
of the power at the exit of the sample is measured. Our experimental
setup used a cw laser ($\lambda=532\,\nano\meter$, Verdi, Coherent), and the
power of the laser was about $30\,\milli\watt$ close to the N-I transition and
reaching $100\,\milli\watt$ far from the transition. The beam waist at focus
was about $26\,{\mu}$m, and an osciloscope (Tektronics, Standfor)
was used for data acquisition. The polarization ($\mathbf{E}$) of
the laser beam was set either parallel or perpendicular to the nematic
director ($\mathbf{n}$). For these geometrical configurations the
optical torque is null, i.~e., director reorientation is not expected.

In systems exhibiting nonlinear absorption, for sufficiently thin samples,
and to first-order corrections in the irradiance, the normalized transmittance
in the open-aperture ($\Gamma_{o}$) configuration of the Z-scan technique
is given by \cite{Chapple}
\begin{equation}
\Gamma_{o}=1-\frac{\Theta}{\left[1+\left(\frac{z}{z_{0}}\right)^{2}\right]}.\label{eq:1}
\end{equation}
Here $z_{0}$ is the Rayleigh range of the beam, $\Theta=\beta I_{0} L_{ef}/2$,
where $L_{ef}=\left[1-\exp\left(-\alpha_{0}L\right)\right]/\alpha_{0}$
is the effective thickness of the sample, $\beta$ is the nonlinear
optical absorption coefficient, and\emph{\ }$I_{0}$ is the irradiance
at the waist of the laser beam. Although Eq.~\ref{eq:1} is derived
assuming a two-photon absorption process, it can be applied to fit
the open-aperture Z-scan curve in any case, assigning to $\beta$ the
meaning of an effective nonlinear optical absorption coefficient.

\section{Experimental results and discussions}

In Figures 2(a) and 2(b), we show typical open-aperture $Z$-scan
traces, obtained at $T=35\,\celsius$ (nematic phase) for the two
relative orientations of the nematic director and beam polarization,
$\mathbf{n}\parallel\mathbf{E}$ and $\mathbf{n}\perp\mathbf{E}$,
respectively. The error bars correspond to the standard error of the
mean for at least ten measurements in each $z$ position. As can be
seen, for $\mathbf{n}\parallel\mathbf{E}$ the transmittance of the
sample increases when approaching the focal point, showing that the
nonlinear optical absorption is negative ($\beta<0$). For
$\mathbf{n}\perp\mathbf{E}$ the transmittance decreases as the focal
point is approached, which shows that the nonlinear optical absorption
is poistive ($\beta>0$). Whitin the sensitivity of our experimental
setup, no nonlinear optical absorption was observed in the isotropic
phase ($\beta_{iso}\simeq0$). We have also checked the nonlinear
optical response of the empty glass cell with PVA coating: for the
intensities in our experiment the cell does not show any
nonlinear optical response.

In Figure 3 we show $\beta$ in the nematic region and in the isotropic
phase, for polarization of the incident beam parallel and
perpendicular to the director. For each relative orientation of the
the director and the optical field, the $\beta$'s resulting from the
fittings were rather independent of the laser beam power. For
$\mathbf{n}\parallel\mathbf{E}$, the sample displays $\beta<0$ and for
$\mathbf{n}\perp\mathbf{E}$, the sample displays $\beta>0$ over the entire
nematic region, i.~e., $\beta_{||}<0$ and $\beta_{\perp}>0$, with increasing
absolute values as the N-I transition temperature is approached
from below. It is also clear that the magnitude of $\beta_{\parallel}$ is
bigger than that of $\beta_{\perp}$. The fitting of the experimental
data with Eq.~\ref{eq:1} gives an effective nonlinear absorption
coefficient $\beta$ for the nonlinear optical process observed in the
E7 liquid crystal of about $-10^{-2}\,\milli\per\watt$ and
$10^{-3}\,\milli\per\watt$, for $\beta_{\parallel}$ and $\beta_{\perp}$,
respectively.

The origin of the nonlinear optical response in the nematic phase and
its enhancement when aproaching the N-I transition is not clear
yet. The optical absorption of liquid crystals containing satured and
unsatured phenyl is relevant in the UV spectral region, involving
$\pi\rightarrow\pi^{*}$ electronic transitions, which have negligible
absorption in the visible range \cite{Khoo}.  As shown by the
pioneering work of Chatelain \cite{Chatelain,Degennes}, the linear
scattering due to fluctuations of the nematic director is strong for
small scattering wavevector $\mathbf{q}$ and weakly dependent on
temperature. Measurements of the scattering in a E7 liquid-crystal
sample showed an approximately constant scattering coefficient over
almost all of the nematic range, with a light increase in a
temperature interval of about $5\,\celsius$ in the vicinity of
$T_{NI}$ \cite{Wu-2}. 
Linear scattering makes the transmitted power independent of the
sample position, leading to a flat response in the normalized
transmittance curve in an open-aperture Z-scan measurement. Another source of many nonresonant stimulated scattering processes, such as
Stimulated Raman Scattering (SRS), Stimulated Brillouin Scattering
(SBS), Stimulated Orientational Scattering (SOS) and Stimulated
Thermal Scattering (STS), have been observed in thermotropic liquid
crystals under appropriate conditions \cite{Khoo5}.  Usually the SRS and
SBS processes require long interation paths (of the order of
1\,\centi\meter) and very strong lasers \cite{Etchegoin-1,Narasimha}. 
In the case of SOS, the forward orthogonal polarized beam generated by
stimulated scattering grows like
$I(z)=I_{0}^{\mathrm{noise}}e^{\left(G_{o}I_{e}-\alpha\right)z}$, where $G_{o}$
is the intensity gain factor and $\alpha$ is an effective loss
coefficient combining linear absorption and scattering processes. For a
$200\,\micro\meter$-thick sample of E7 liquid crystal, the threshold
intensity is $\sim 10^{3}\,\watt\per\centi\meter\squared$ \cite{khoo4},
comparable to the intensity in our setup for
$\mathbf{n}\perp\mathbf{E}$. For a $20\,\micro\meter$-thick sample, therefore,
the threshold intensity for an SOS process is one order of magnitude 
higher. Finally, a highly absorbing medium is necessary to trigger a STS
process. In a $150\,\micro\meter$ planar aligned dye-doped sample
of 5CB, a STS process is observed when the interaction length is about
$1\, \milli\meter$, with oblique incidence and an incident laser
power ($\lambda=514.5\,\nano\meter$) of $\sim400
\,\milli\watt$ \cite{Khoo-6}. Any nonlinear scattering in an open-aperture Z-scan experiment should
yield results resembling nonlinear absorption with $\beta>0$, because along with the
dominant forward beam this process originates a small amount of
backscattered radiation for both incident lasers $I_{0}$ and $I_{e}$
($\mathbf{n}\perp\mathbf{E}$ and $\mathbf{n}\parallel\mathbf{E}$
respectively). Studies of the nonlinear scattering show that it is
also nearly temperature independent \cite{Khoo3} and inversely
proportional to the square of the beam wavelength, i.~e., proportional
to $\lambda^{-2}$ \cite{Khoo2}. A similar study of a
planar oriented sample of 5CB \cite{Li}, using the
$\lambda=514.5\,\nano\meter$ line of an argon laser at comparable
power, showed no nonlinear optical absorption, either in the nematic
phase or close to the N-I transition. 

Although the wavelength in our setup lies close to the maximum
absorbance of the E7 liquid crystal, two-photon absorption can be
ruled out in view of the low laser power. Two-photon absorption with
low-power lasers has only been observed in highly absorbing dye-doped
liquid crystals, which is not the case here. A more likely source of
nonlinear absorption is laser-induced reorientation of absorbing species as seen
in a $50\,\micro\meter$-thick homeotropically oriented sample of
dye-doped 5CB \cite{Tabiryan}. An optical field exerts an optical
torque proportional to the dieletric anisotropy $\Delta\varepsilon$ \cite{Khoo2}. From the balance between the elastic torque and the
optical torque, under hard-boundary and small director-reorientation
conditions, it can be shown that the maximum reorientation
$\theta_{max}$ for an extradordinary ray is proportional to
$\Delta\varepsilon/K_{1}$, where $K_{1}$ is the elastic constant for
splay distortion. Since $\Delta\varepsilon\propto S$ and $K_{1}\propto
S^{2}$, the maximum reorientation in the nematic phase satisfies
$\theta_{max}\propto S^{-1}$. It follows that $\theta_{max}$ increases
as $T$ approaches $T_{NI}$. In the case of an ordinary ray, the
threshold intensity for diretor reorientation is proportional do
$K_{3}/\Delta\varepsilon\propto S$ where $K_{3}$ is the elastic
constant for bend distortion. Hence, the nonlinear absorption observed
in the E7 liquid crystal below $T_{NI}$ could be due to the
reorientation of absorbing species present in the liquid cristalline
matrix along with a reduction of ordening associated with heating that
leads to absorption depletion, a reduction expected to be more
effective in the case of an extraordinary ray.

In the nematic phase, given the dichroic ratio
$R=\alpha_{\parallel}/\alpha_{\perp}$ of the sample, i. e., the ratio
between the linear absorbances of radiation with the polarization
parallel and perpendicular to director, an optical order parameter
$S=(R-1)/(R+2)$ can be defined, provided the transition moment is
parallel to the molecular axis \cite{Korte}. By analogy, we can define
a nonlinear dichroic ratio
$R_{NL}=|{\beta_{\parallel}}/{\beta_{\perp}}|$ and the parameter
$S_{NL}=(R_{NL}-1)/(R_{NL}+2)$. Figure 4 displays $S_{NL}$ as a
function of $T$. Clearly, $S_{NL}$ vanishes at the N-I transition and
is equal to zero in the isotropic phase. 

Figure 5 displays $S_{NL}$ as a function of the reduced temperature in
log scales.  Also shown is a linear fit, close the transition
temperature, under the assumption that
$S_{NL}\propto\left(1-\frac{T}{T^{\dagger}}\right)^{\phi}$ , where
$\phi$ is the critical exponent and $T^{\dagger}$ is the temperature
of the effective or virtual second order transition seen from below
$T_{NI}$, which is different from the spinodal temperature
$T^{^{\star\star}}$, the absolute limit of the nematic phase upon
heating.  For an homologus series of liquid crystals it has been shown
that $T^{\dagger}-T_{NI}\simeq0.2\,\celsius$ \cite{Chirtoc}, which is
of the order of the experimental uncertainty in our measurements.
This considered and attention paid to the sensitivity of the fitting to
the choice of $T^{\dagger}$, we proceded to vary $T^{\dagger}$
between $58.1\,\celsius$ and $58.4\,\celsius$. The resulting average
critical exponent $\phi$ was $0.22\pm0.05$, close to the
accepted value for the order-parameter critical exponent under the
assumption that the N-I transition has tricritical character.

\section{Conclusions}

We have investigated the nonlinear optical absorption of an
homogeneously oriented sample of liquid crystal E7 in the nematic
phase and close to the N-I transition under continuos $532\,
\nano\meter$ wave excitation.  The effective nonlinear optical
absorption exhibits opposite character for the two geometrical
configurations of the nematic director relative to the polarization of
the incident beam, i.~e., $\beta_{||}<0$ and $\beta_{\perp}>0$. We
tentatively ascribe this behavior to the combination of an
optical-field induced reorientation of the nematic director and
heating of the sample due to absorbing species in the liquid
cristalline matrix. Interestingly, the coefficiente $S_{NL}$, defined
from the $\beta$'s in an analogy to the optical order parameter
\emph{S } defined from the linear optical-absorption coefficients,
satisfies an order-parameter power law with critical exponent
$0.22\pm0.05$, which points to a N-I tansition with tricritical
character.

\bigskip{}

\textbf{Acknowledgments}

This work had the financial support of the Brazilian agencies CAPES,
CNPq, FAPESP, Funda\c c\~ao Arauc\'aria and the Secretaria de Ci\^encia, Tecnologia
e Ensino Superior do Governo do Paran\'a, and it was conducted as part
of the research program of the Instituto Nacional de Ci\^encia e Tecnologia
de Fluidos Complexos (INCT-FCx).

\bigskip{}

\begin{center}
\textbf{Figure Captions} 
\end{center}

Fig. 1: Sketch of the $Z$-scan apparatus in the closed-aperture configuration:
The lens, sample, iris, and photodetector are labeled L, S, I, and PD, respectively.
\\

Fig. 2: Typical $Z$-scan curves in open-aperture configuration at
$T=35\,\celsius$: (a) $\mathbf{n}\parallel\mathbf{E}$ for incident
power $P=3\,\milli\watt$; (b) $\mathbf{n}\perp\mathbf{E}$ for incident
power $P=40\,\milli\watt$. Solid lines are theoretical fits using Eq.~\ref{eq:1}.
\\

Fig. 3: Nonlinear optical absorption coefficient $\beta$ as a function of temperature
in the nematic range, for both configurations of the nematic director
relative to the beam polarization. The triangles ($\blacktriangle$)
and circles ($\circ$) correspond to $\mathbf{n}\parallel\mathbf{E}$
and $\mathbf{n}\perp\mathbf{E}$, respectively. The dotted vertical line
indicates the clearing temperature.
\\

Fig. 4: Coefficient $S_{NL}$ as a function of temperature.
The dotted line indicates the clearing temperature.
\\

Fig. 5. Log-log plot of the absolute value of $S_{NL}$ as a function
of the reduced temperature. The solid line represents a typical linear
fit to the data closest to the transition temperature. The dotted line
indicates the clearing temperature.\\

\begin{figure}[htb]
\includegraphics[scale=0.3]{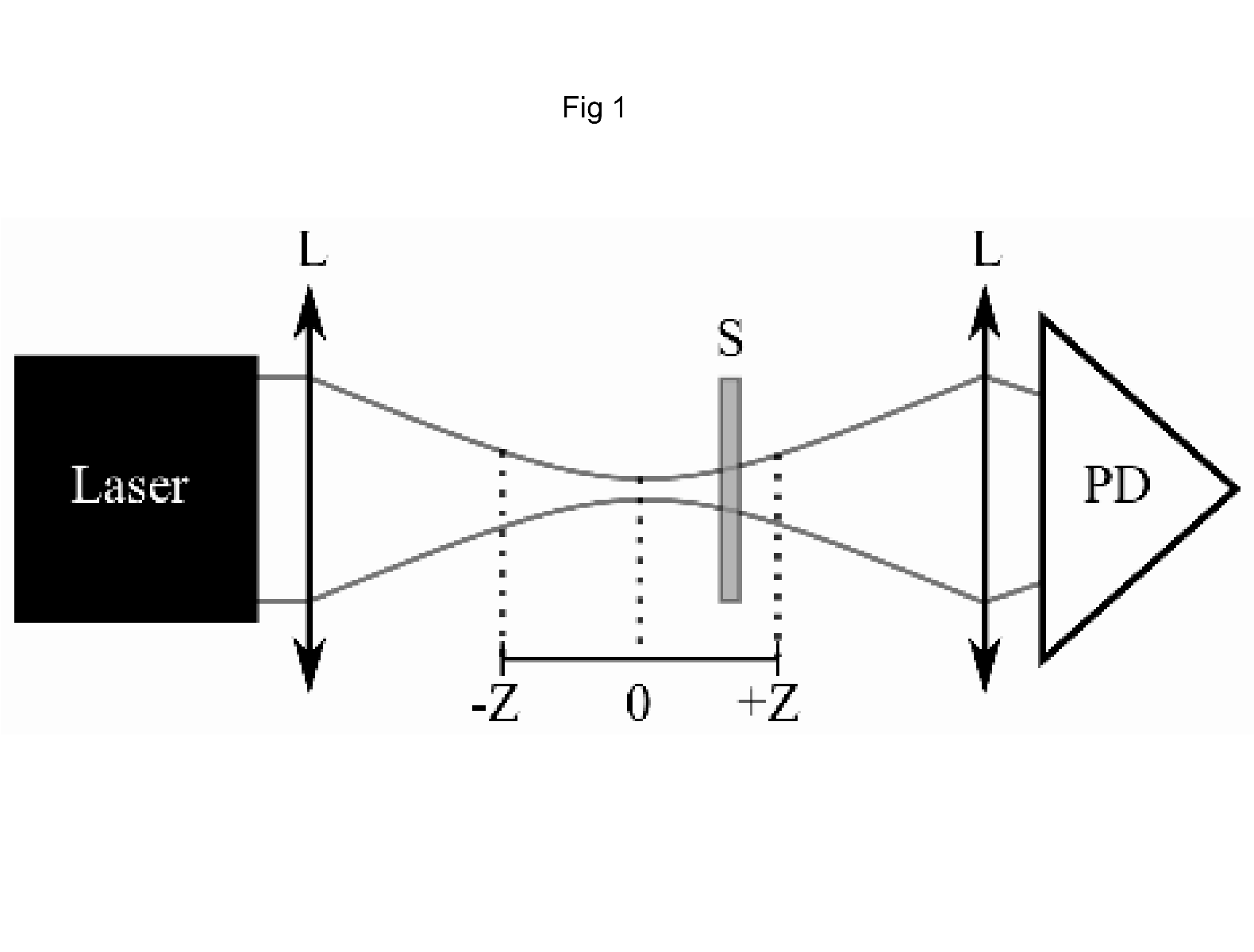}
\end{figure}

\begin{figure}
\includegraphics[scale=0.7]{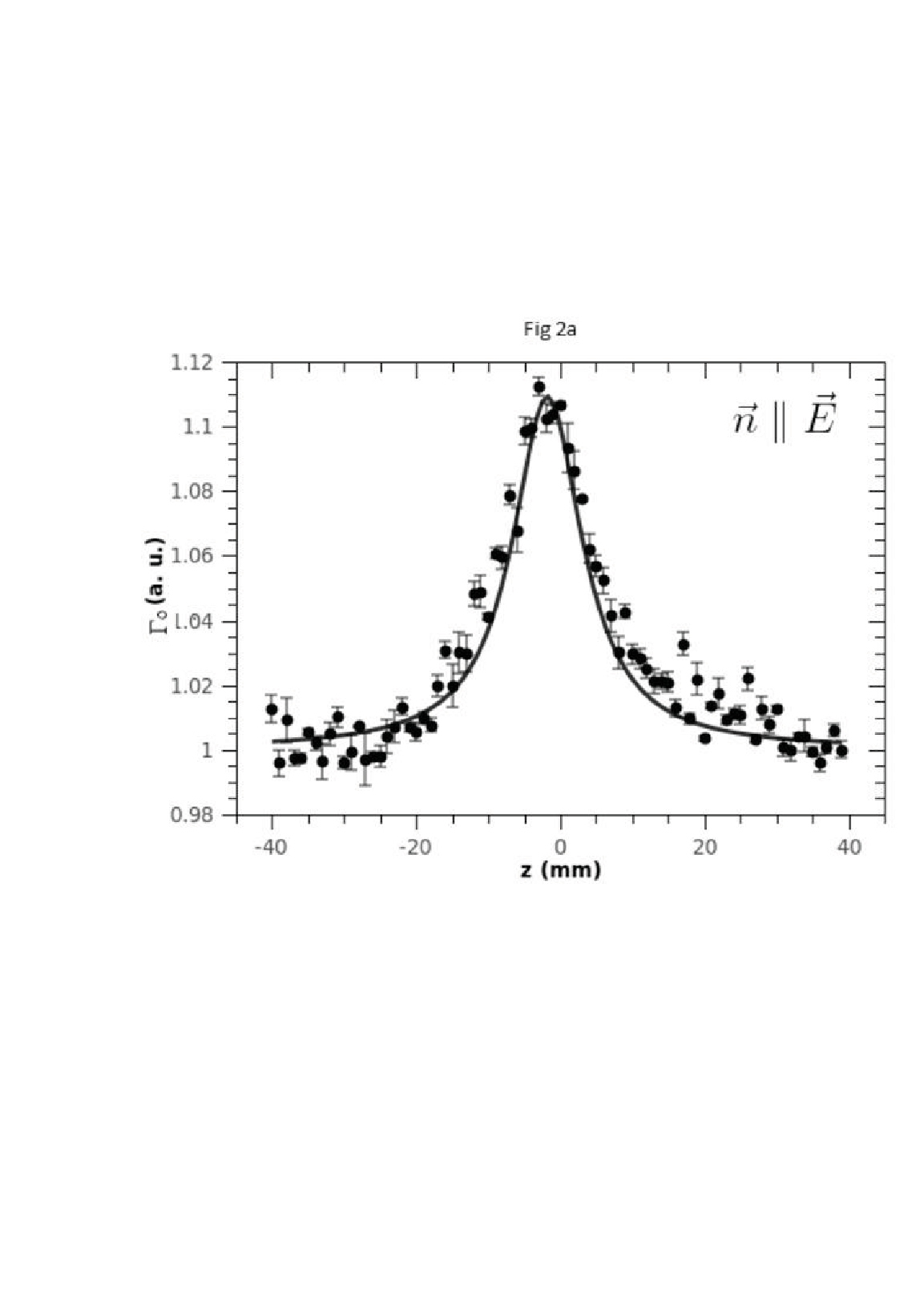}
\end{figure}

\begin{figure}
\includegraphics[scale=0.7]{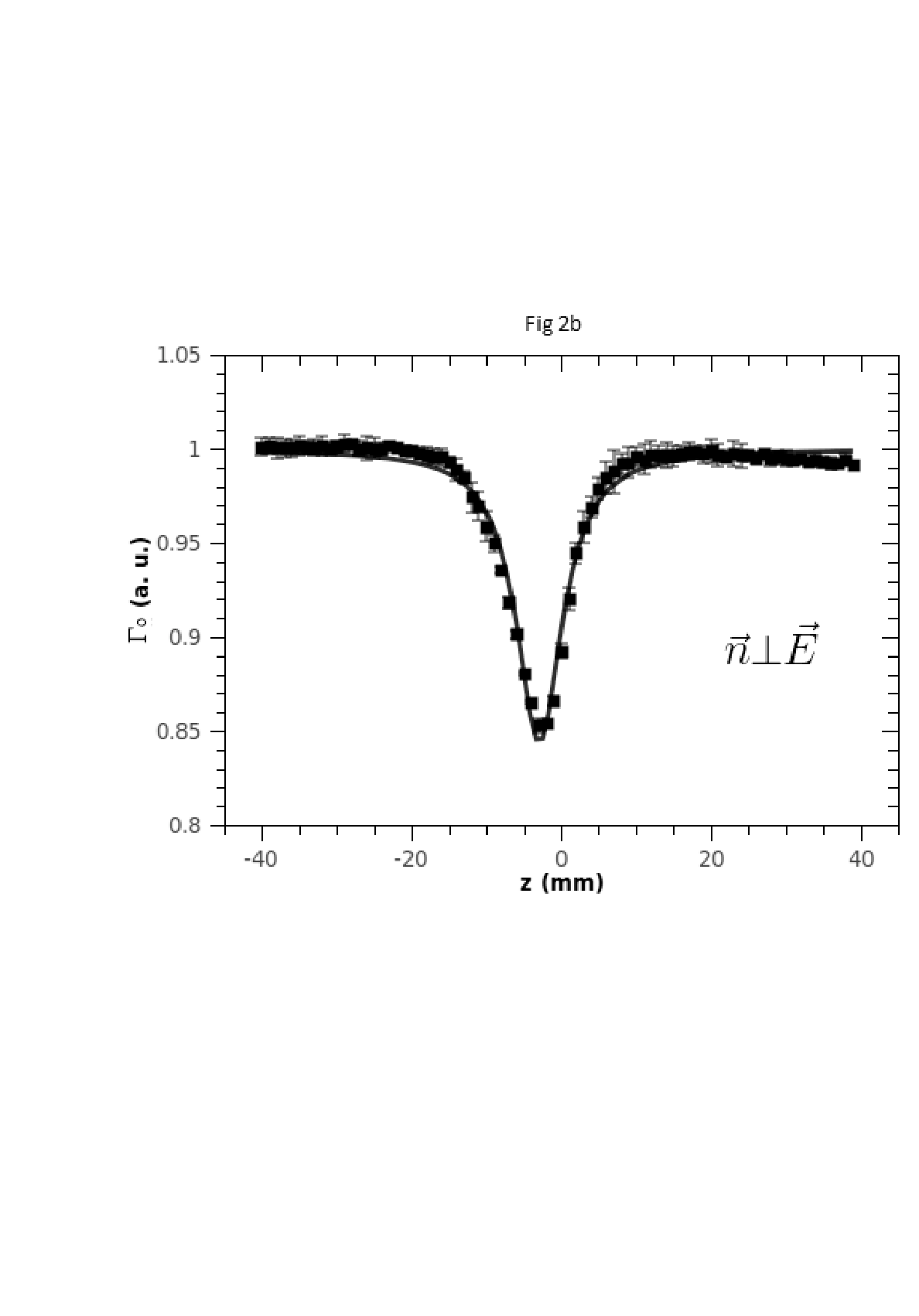}
\end{figure}

\begin{figure}
\includegraphics[scale=1.5]{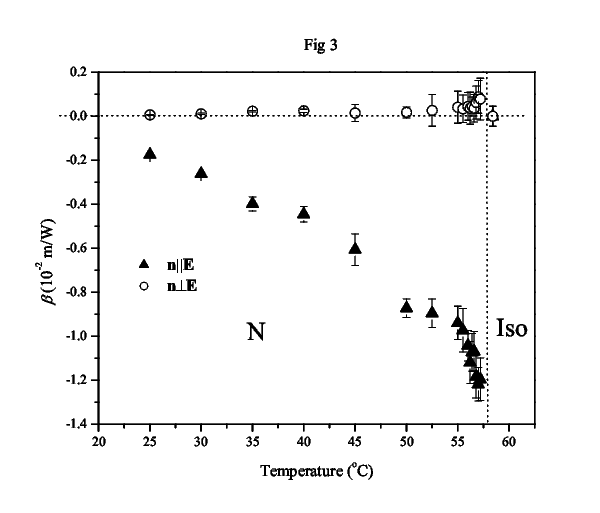}
\end{figure}

\begin{figure}
\includegraphics[scale=1.5]{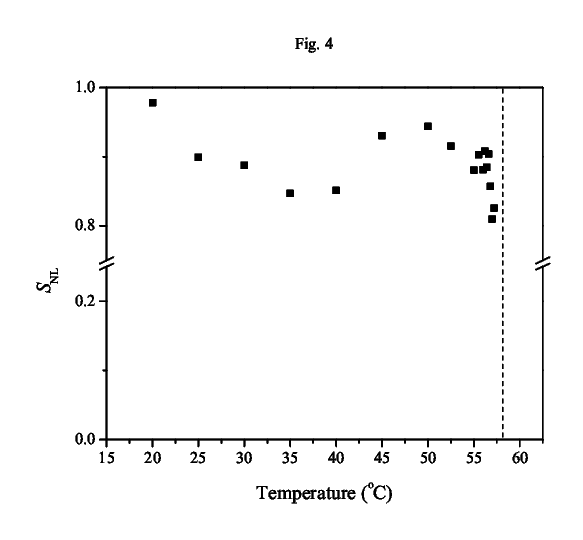}
\end{figure}

\begin{figure}
\includegraphics[scale=1.5]{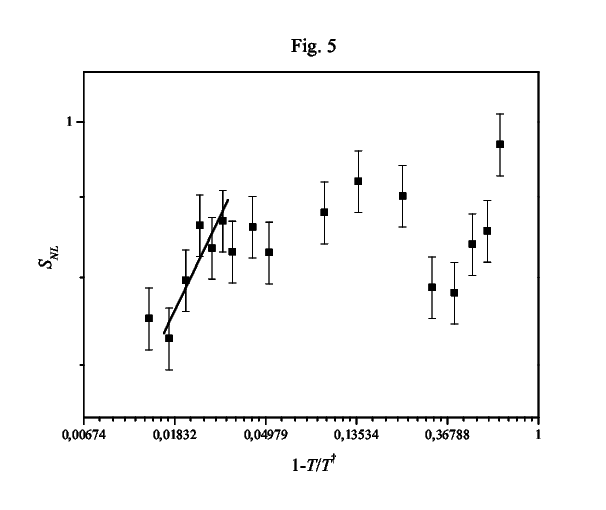}
\end{figure}

\end{document}